\documentclass{elsart}

\newcommand{\bm}[1]{\mbox{\boldmath $ #1 $}}
\usepackage[dvips]{graphicx}
\begin{document}
\runauthor{Matsuzaki and Tanigawa}
\begin{frontmatter}
\title{Phenomenological construction of a relativistic nucleon-nucleon interaction 
for the superfluid gap equation in finite density systems}
\author[fue]{\underline{Masayuki Matsuzaki}\thanksref{emm}}
\author[kyushu]{Tomonori Tanigawa\thanksref{tmm}}

\address[fue]{Department of Physics, Fukuoka University of Education, 
Munakata, \\
Fukuoka 811-4192, Japan}
\address[kyushu]{Department of Physics, Kyushu University, 
Fukuoka 812-8581, Japan}
\thanks[emm]{Corresponding author, Electronic address: matsuza@fukuoka-edu.ac.jp, \\
~~~Tel: +81-940-35-1377, Fax: +81-940-35-1712}
\thanks[tmm]{Electronic address: tomo2scp@mbox.nc.kyushu-u.ac.jp}
\begin{abstract}
We construct phenomenologically a relativistic particle-particle channel 
interaction which suits the gap equation for 
nuclear matter. This is done by introducing a {\it density-independent} 
momentum-cutoff parameter to the relativistic mean field 
(Hartree and Hartree-Fock) models 
so as to reproduce the pairing properties obtained by the Bonn-B 
potential and not to change the saturation property. 
The interaction so obtained can be used for the Relativistic 
Hartree-Bogoliubov calculation, but some reservation is necessary 
for the Relativistic Hartree-Fock-Bogoliubov calculation.
\end{abstract}
\end{frontmatter}
\newpage

\section{Introduction}

  Pairing correlation between nucleons is a key ingredient 
to describe the structure of neutron stars and finite nuclei. 
There are two distinct ways of description of such {\it finite-density} 
nuclear many-body systems; the non-relativistic and the relativistic ones. 
The latter incorporates the mesons 
explicitly in addition to the nucleons in terms of a field theory. 
Both describe the basic properties such as the saturation with a similar 
quality in different manners. 
Irrespective of whether non-relativistic or relativistic, however, 
various theoretical approaches can be classified into two types: 
One is realistic studies adopting phenomenological interactions 
constructed for finite-density systems from the beginning (hereafter 
we call this the P-type, indicating ``phenomenological"), as often 
done in the studies of heavy nuclei. And the other is microscopic 
studies based on bare nucleon-nucleon interactions in free space 
(the B-type, indicating ``bare"). 
In relativistic studies, typical examples of these two types as for the 
particle-hole (p-h) channel are the relativistic mean field 
(RMF) model and the Dirac-Brueckner-Hartree-Fock (DBHF) method, 
respectively. As for the particle-particle (p-p) channel, that is, 
pairing correlation, a bare interaction was used as the lowest-order 
contribution in the gap equation~\cite{mig} in a study of the 
B-type~\cite{rel3}. This is thought to be a good approximation at least 
for the $^1S_0$ channel (Refs.\cite{baldo,gar}, for example).
The first relativistic study of the P-type of pairing correlation 
in nuclear matter was done by Kucharek and Ring~\cite{kr}. 
They adopted a one-boson exchange (OBE) interaction with the coupling 
constants of the RMF model, which we call the RMF interaction hereafter, 
aiming at a fully selfconsistent Hartree-Bogoliubov calculation, 
which we call the P1-type, in the sense that both the p-h and the p-p 
channel interactions are derived from a common Lagrangian. 
But the resulting pairing gaps were about three times larger than 
those accepted in the non-relativistic studies (see Fig.\ref{f6}). 
The reason can be ascribed to the fact that the coupling constants of the 
RMF model were determined by physics involving only low momenta 
($k\le k_{\rm F}^0$, $\frac{2}{3\pi^2}(k_{\rm F}^0)^3=\rho_0$ 
denoting the saturation density of symmetric nuclear matter), 
and therefore the adopted OBE interaction is not reliable at 
high momenta. After a five-year blank, some attempts to improve this were 
done~\cite{rel1,rel2,mm}. But their results were insufficient. 

  An alternative way is to adopt another interaction in the p-p channel 
while the single-particle states are still given by the RMF model. 
We call this the P2-type. There are some variations of this. 
The first one, which we call the P2a-type, adopts another phenomenological 
interaction for the p-p channel. Actually, the non-relativistic Gogny 
force~\cite{gogny} was used combined with the single-particle states 
of the RMF model for finite nuclei in Ref.\cite{rhfb} and subsequent works, 
and gave excellent results. The second variation, which we call the P2b-type, 
is to adopt a bare interaction that describes the high-momentum part 
realistically, the Bonn potential, again combined with 
the single-particle states of the RMF model~\cite{rr} (see also Ref.\cite{mt}). 
If one assumes that the RMF model simulates roughly the DBHF calculation, 
this P2b-type can be regarded as simulating the B-type calculation~\cite{rel3} 
mentioned above. The results 
of these P2a- and P2b-types are very similar at densities $\rho<\rho_0$. 
This supports a statement that the Gogny force resembles a realistic free 
interaction in the low-density limit~\cite{bert} (see also Ref.\cite{gar}). 
But a clear difference 
can be seen at $\rho\sim\rho_0$. This difference can also be seen in 
fully non-relativistic calculations; compare the results in 
Refs.\cite{nu1} and~\cite{nu2}, for example. The precise origin of this 
difference has not been understood well. A comparison after taking 
the polarization effects which have been known to be important at 
finite densities~\cite{chen,pol1,pol2} into account may be necessary 
(see also Ref.\cite{mt}).
The third variation, which we call the P2b$^\prime$-type, is to parameterize 
the p-p channel interaction in terms of a few scattering 
parameters. This was actually examined by being combined with the DBHF 
calculation~\cite{shift}, which we call the B$^\prime$-type (see also 
Ref.\cite{papen}). 
This method is free from model-dependent ambiguities and meets the 
viewpoint of modern effective field theories~\cite{wein2,sw2} but is 
applicable only to dilute systems. These classifications are summarized 
in Table 1.

  Since we are interested in a wide density range where the $^1S_0$ 
gap exists and would like to respect the selfconsistency in the sense 
that both the p-h and the p-p channel interactions are derived from a 
common Lagrangian, here we construct a phenomenological relativistic 
nucleon-nucleon interaction based on the RMF interaction (the P1-type) 
by adjusting to the pairing properties given by the RMF+Bonn calculation 
(the P2b-type). In other words, we aim at constructing an interaction 
similar to the Gogny force in the sense that it reproduces the pairing 
properties given by the bare interactions in spite of the fact that 
it was constructed for the finite-density system from the beginning. 

  The contents of this paper are as follows: In sect. 2 we present our 
method of constructing a phenomenological p-p channel 
interaction for the superfluid gap equation. This procedure is applied 
both to the Hartree and to the Hartree-Fock model. In sect. 3, we discuss 
the pairing properties obtained in the Hartree approximation. Note here 
that we adopt the no-sea approximation throughout this paper and therefore 
the Relativistic Hartree model means the so-called Relativistic Mean 
Field model. In sect. 4, the Relativistic Hartree-Fock-Bogoliubov 
calculations with or without the modulation of the high-momentum interaction 
are presented. Conclusions are given in sect. 5. 

\section{Construction of a relativistic particle-particle channel interaction}

  We start from the ordinary $\sigma$-$\omega$ model Lagrangian density,
\begin{eqnarray}
{\mathcal L}&=&\bar\psi(i\gamma_\mu\partial^\mu-M)\psi
\nonumber\\
 &+&\frac{1}{2}(\partial_\mu\sigma)(\partial^\mu\sigma)
  -\frac{1}{2}m_\sigma^2\sigma^2
  -\frac{1}{4}\Omega_{\mu\nu}\Omega^{\mu\nu}
  +\frac{1}{2}m_\omega^2\omega_\mu\omega^\mu
\nonumber\\
 &+&g_\sigma\bar\psi\sigma\psi-g_\omega\bar\psi\gamma_\mu\omega^\mu\psi ,
\nonumber\\
&\Omega_{\mu\nu}&=\partial_\mu\omega_\nu-\partial_\nu\omega_\mu .
\end{eqnarray}
The antisymmetrized matrix element of the RMF interaction $V$ derived from 
this Lagrangian is defined by, 
\begin{equation}
\bar v({\bf p},{\bf k})
=\langle{\bf p}s',\tilde{{\bf p}s'}\vert V\vert
           {\bf k}s,\tilde{{\bf k}s}\rangle
   -\langle{\bf p}s',\tilde{{\bf p}s'}\vert V\vert
           \tilde{{\bf k}s},{\bf k}s\rangle ,
\end{equation}
under an instantaneous approximation, with tildes denoting time reversal. 
After a spin average and an angle integration are performed to project out 
the $S$-wave component, its concrete form is given by
\begin{eqnarray}
&&\bar v(p,k)
\nonumber \\
&&=-\frac{g_\sigma^2}{2E^\ast_pE^\ast_k}
          \Biggl\{1+\frac{2(E^\ast_pE^\ast_k+M^{\ast\,2})-(p^2+k^2+m_\sigma^2)}{4pk}
                    \ln{\left(\frac{(p+k)^2+m_\sigma^2}{(p-k)^2+m_\sigma^2}\right)}
          \Biggr\}
\nonumber \\
&&\,+\frac{g_\omega^2}{2E^\ast_pE^\ast_kpk}\left(2E^\ast_pE^\ast_k-M^{\ast\,2}\right)
              \ln{\left(\frac{(p+k)^2+m_\omega^2}{(p-k)^2+m_\omega^2}\right)}\, ,
\nonumber
\end{eqnarray}
where
\begin{equation}
E^\ast_k=\sqrt{{\bf k}^2+M^{\ast\,2}}\, .
\end{equation}
  Our policy of constructing a phenomenological interaction proposed above is to 
introduce a {\it density-independent} parameter $\Lambda$ so as not to change 
the Hartree part with the momentum transfer ${\bf q}=0$ which determines 
the single-particle energies, 
respecting that the original parameters of the RMF are density-independent. 
In the Hartree-Fock model, the ${\bf q}\neq0$ part also contributes to the 
mean field. This will be discussed in sect.4 in detail. 
Since the high-momentum part of the RMF interaction does not have 
a firm experimental basis as mentioned above, we suppose there is room 
to modify that part. Needless to say, such a modification should be checked 
by studying independent phenomena, for example, medium-energy 
heavy-ion collisions. 
Some adjustments of the density-dependent $\delta$ force to the ones that 
give realistic pairing properties have already been examined in the 
non-relativistic studies \cite{bert,gar}. Among them, a fit to a bare 
interaction in the $T=0$ channel was done \cite{gar}. We aim at a similar 
procedure in the relativistic model for the $T=1$ channel. 
In the preceding letter~\cite{tm}, the upper bounds of the momentum 
integration in the gap equation, 
\begin{equation}
\Delta(p)
=-\frac{1}{8\pi^2}
  \int_0^\infty \bar v(p,k)
         \frac{\Delta(k)}{\sqrt{(E_k-E_{k_{\rm F}})^2+\Delta^2(k)}}k^2dk\, ,
\label{gapeq}
\end{equation}
where
\begin{equation}
E_k=E^\ast_k+g_\omega\langle\omega^0\rangle \, ,
\label{spe}
\end{equation}
and the nucleon effective mass equation, 
\begin{equation}
  M^\ast=M-\frac{g_\sigma^2}{m_\sigma^2}\frac{\gamma}{2\pi^2}
  \int_0^\infty 
  \frac{M^\ast}{\sqrt{k^2+M^{\ast\,2}}}v_k^2k^2dk\, ,
\label{effmass}
\end{equation}
where the spin-isospin factor $\gamma$ = 4 and 2 indicate 
symmetric nuclear matter and pure neutron matter, respectively, were cut 
at a finite value $\Lambda$, as usually done in condensed-matter physics, 
since the gap increases monotonically until reaching Kucharek and 
Ring's value when the model space is enlarged as shown in Fig.3 of 
Ref.\cite{mm}, while $\bar v(p,k)$ is left unchanged. We call this method the 
sudden cutoff hereafter. This was done first in Ref.\cite{kr} by inspection. 
We proposed a quantitative method to determine $\Lambda$, which is described 
below, and obtained 3.60 fm$^{-1}$ for the linear $\sigma$-$\omega$ 
parameter set in Ref.\cite{tm}. This value almost coincides with their value, 
about 3.65 fm$^{-1}$, for the NL2 parameter set. 

  In the present paper, we examine smooth cutoffs that weaken the 
high-momentum part; a form factor $f({\bf q}^2)$, ${\bf q}={\bf p}-{\bf k}$, 
is applied to each nucleon-meson vertex in 
$\bar v({\bf p},{\bf k})$ while the upper bounds of 
the integrals are conceptually infinity. They are replaced numerically by 
a finite number, 20 fm$^{-1}$ which has been proved to be large enough in 
Ref.\cite{mm}. Since there is no decisive reasoning to choose 
a specific form, we examine four types,
\begin{eqnarray}
\mbox{monopole: }\,\,f({\bf q}^2)&=&\frac{\Lambda^2}{\Lambda^2+{\bf q}^2}\, ,
\nonumber \\
\mbox{dipole: }\,\,f({\bf q}^2)&=&\left(\frac{\Lambda^2}
                                             {\Lambda^2+{\bf q}^2}\right)^2\, ,
\nonumber \\
\mbox{strong(a): }\,\,f({\bf q}^2)&=&\frac{\Lambda^2-{\bf q}^2}
                                       {\Lambda^2+{\bf q}^2}\, .
\nonumber \\
\mbox{strong(b): }\,\,f({\bf q}^2)&=&\frac{\Lambda^2-m_i^2}
                                       {\Lambda^2+{\bf q}^2}\, , \quad\quad 
(i=\sigma, \omega) .
\end{eqnarray}
Note that the sudden cutoff above was applied to ${\bf k}$, not to ${\bf q}$.

  The parameter $\Lambda$ is determined so as to minimize the difference 
in the pairing properties from the results of the P2b-type RMF+Bonn 
calculation. Assuming the P2b-type roughly simulates the B-type as mentioned 
in sect.1, conceptually we aim at fitting to the pairing properties given 
by the fully microscopic B-type calculation. 
Here we adopt the Bonn-B potential because this has a moderate property 
among the available (charge-independent) versions A, B, and C~\cite{bonn}. 
The pair wave function, 
\begin{eqnarray}
&&\phi(k)=\frac{1}{2}\frac{\Delta(k)}{E_{\rm qp}(k)}\, , 
\nonumber \\
&\,&E_{\rm qp}(k)=\sqrt{(E_k-E_{k_{\rm F}})^2+\Delta^2(k)}\, ,
\label{pairwf}
\end{eqnarray}
is related to the gap at the Fermi surface,
\begin{equation}
  \Delta(k_{\rm F})=-\frac{1}{4\pi^2}
  \int_0^\infty \bar v(k_{\rm F},k)\phi(k)k^2dk\, ,
\end{equation}
and its derivative determines the coherence length~\cite{coh}, 
\begin{equation}
\xi=\left(\frac{\int_0^\infty\vert\frac{d\phi}{dk}\vert^2k^2dk}
               {\int_0^\infty\vert\phi\vert^2k^2dk}\right)
      ^\frac{1}{2}\, ,
\end{equation}
which measures the spatial size of the Cooper pairs. These expressions 
indicate that $\Delta(k_{\rm F})$ and $\xi$ carry independent information, 
$\phi$ and $\frac{d\phi}{dk}$, respectively, in strongly-coupled systems, 
whereas they are intimately related to each other in weakly-coupled ones. 
Therefore we search for $\Lambda$ which minimizes
\begin{equation}
\chi^2
=\frac{1}{2N}\sum_{k_{\rm F}}
\left\{\left(\frac{\Delta(k_{\rm F})_{\rm RMF}-\Delta(k_{\rm F})_{\rm Bonn}}
                   {\Delta(k_{\rm F})_{\rm Bonn}}\right)^2
+\left(\frac{\xi_{\rm RMF}-\xi_{\rm Bonn}}
            {\xi_{\rm Bonn}}\right)^2
\right\}\, ,
\end{equation}

where the subscripts ``RMF" and ``Bonn" denote the RMF interaction 
including $\Lambda$ and the Bonn-B potential, respectively, while the 
single-particle states are determined by the original RMF model in both 
cases.

\begin{figure}[t]
\begin{center}
\includegraphics[width=10.32cm,keepaspectratio]{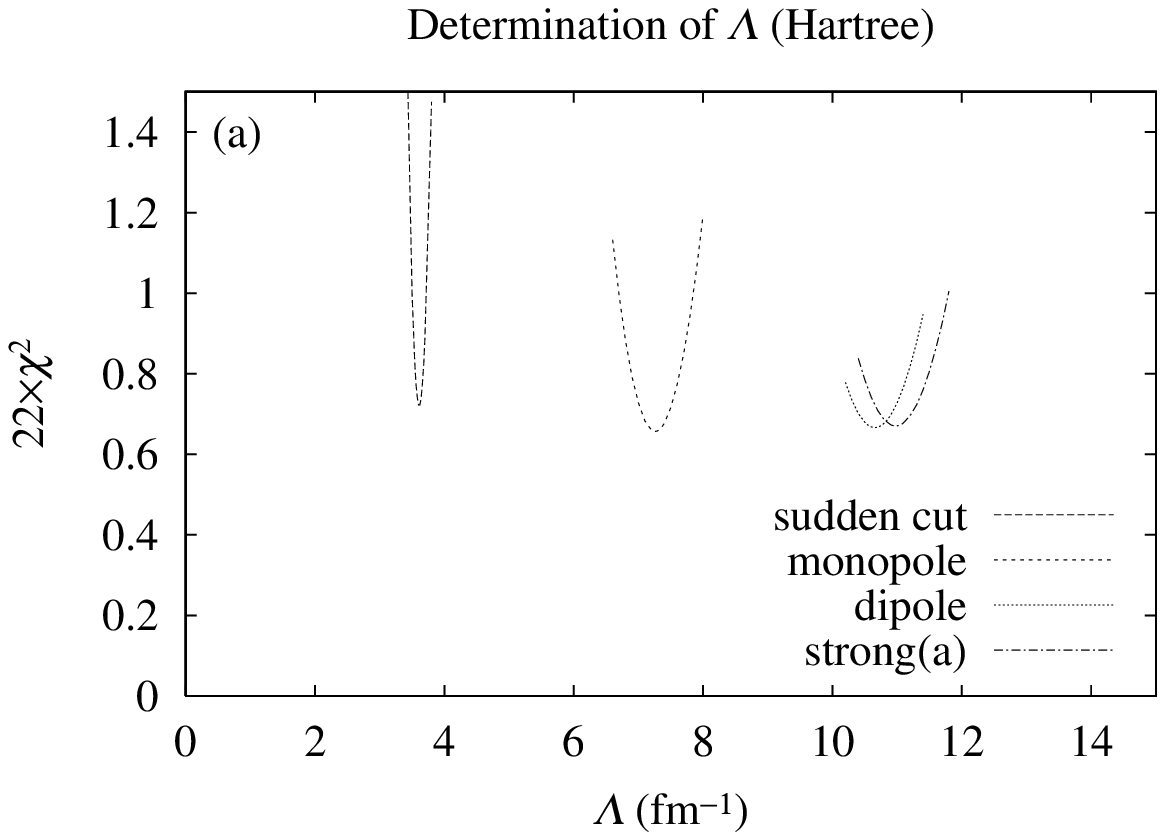}
\vskip 0.5cm
\includegraphics[width=10cm,keepaspectratio]{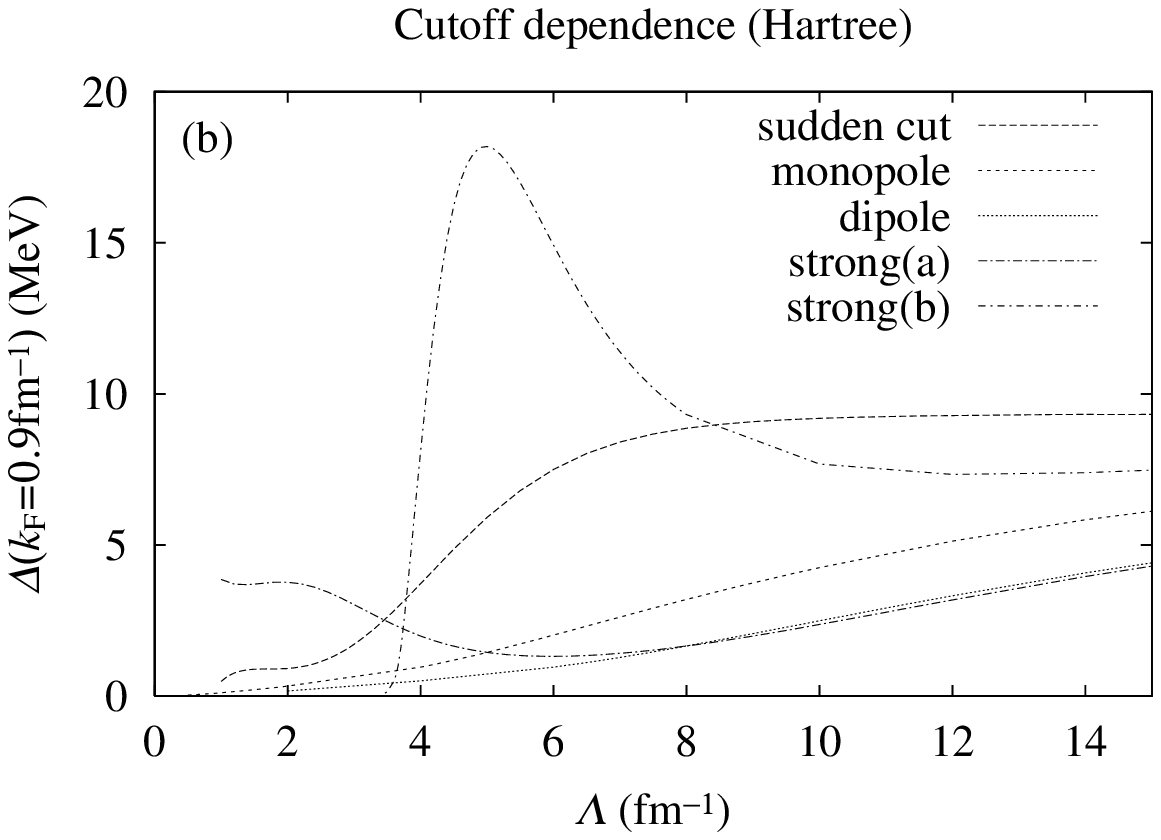}
\end{center}
\caption{
(a): Curvature of $\chi^2$ in Eq.(11) with respect to the 
cutoff parameter $\Lambda$.
(b): $\Lambda$-dependence of the pairing gap at the Fermi 
surface, $k_{\rm F}=$ 0.9 fm$^{-1}$. 
These are the results of the Hartree-Bogoliubov calculation.
}
\label{cutdep}
\end{figure}

  The actual numerical task is to solve the gap equation (\ref{gapeq}) and 
the effective mass equation for the nucleon (\ref{effmass}). 
They couple to each other through Eq.(\ref{spe}) and 
\begin{equation}
v_k^2=\frac{1}{2}\left(1-
   \frac{E_k-E_{k_{\rm F}}}{\sqrt{(E_k-E_{k_{\rm F}})^2+\Delta^2(k)}}\right)\, .
\label{v2}
\end{equation}
The parameters of the standard $\sigma$-$\omega$ model that we adopt are 
$g_\sigma^2=$ 91.64, $g_\omega^2=$ 136.2, $m_\sigma=$ 550 MeV, 
$m_\omega=$ 783 MeV, and $M=$ 939 MeV~\cite{sw}. $N$ in $\chi^2$ is taken 
to be 11; $k_{\rm F}=$ 0.2, 0.3,$\ldots$, 1.2 fm$^{-1}$. In the following, 
the results 
for symmetric nuclear matter are presented. Those for pure neutron matter 
are very similar except that $\Delta(k_{\rm F})$ is a little larger due to 
a larger effective mass $M^\ast$ as shown in Fig.~1\,(b) of Ref.\cite{tm}.
Minimizations of $\chi^2$ give the optimal values, $\Lambda=$ 7.26, 10.66, 
and 10.98 fm$^{-1}$ for the first
three types of form factor, respectively, as shown in Fig.\ref{cutdep}(a). 
We do not choose an optimal $\Lambda$ for the strong(b) type because of its 
pathological $\Lambda$-dependence shown in Fig.\ref{cutdep}(b). 
Figure~\ref{cutdep}(b) shows that the $\Lambda$-dependence of these smooth 
cutoff cases is very mild, except for the strong(b) type, in comparison with 
the sudden cutoff case. The very steep $\Lambda$-dependence in the strong(b) 
case is due to the consecutive depression of the attraction and the 
repulsion at around $\Lambda\sim m_\sigma$ and $\sim m_\omega$, respectively. 
In the strong(a) case, although another smaller $\Lambda$ around 3 fm$^{-1}$ 
can give similar $\Delta(k_{\rm F})$, $\bar v(k_{\rm F},k)$ and consequently 
$\Delta(k)$ exhibit an unphysical staggering at around $k\sim m_\sigma$. 
Therefore we discard this. In addition, although the strong(a) case with 
$\Lambda=$ 10.98 fm$^{-1}$ gives practically the same $k_{\rm F}$-dependence 
of $\Delta(k_{\rm F})$ and $\xi$ (in Fig.\ref{kfdep} shown later), 
$\bar v(p,k)=0$ at $\Lambda=\vert{\bf q}\vert=\vert{\bf p}-{\bf k}\vert$ 
brings about an unphysical staggering in $\Delta(k)$ and $\phi(k)$; this leads 
to an oscillatory structure in $r$-space with a period $\sim \pi/\Lambda$. 
Therefore we discard this, too. 
Form factors with similar $\Lambda$ are also suggested in a study of medium-energy 
heavy-ion collisions~\cite{hi2}. This indicates that the present results 
have a physical meaning. 

\section{Pairing properties obtained in the Relativistic Hartree-Bogoliubov 
calculation by using the constructed interaction}

\begin{figure}[t]
\begin{center}
\includegraphics[width=9.76cm,keepaspectratio]{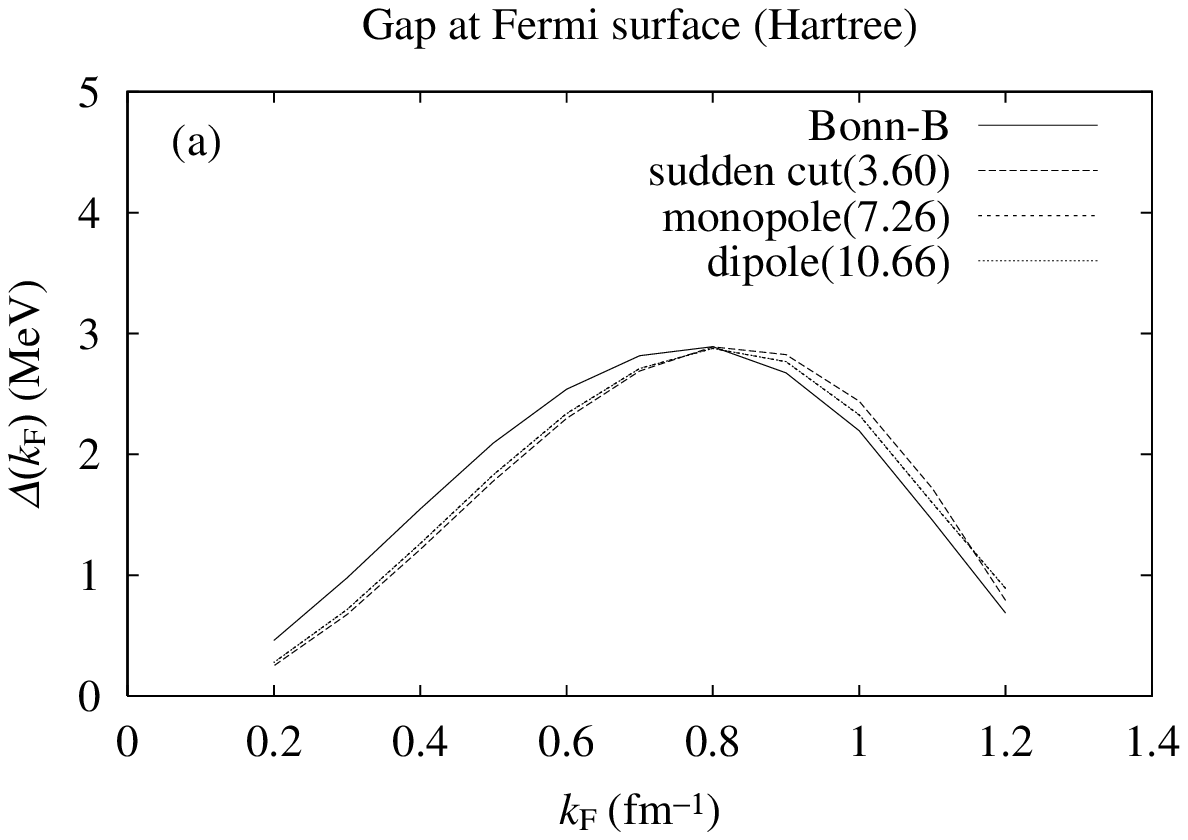}
\vskip 0.5cm
\includegraphics[width=10cm,keepaspectratio]{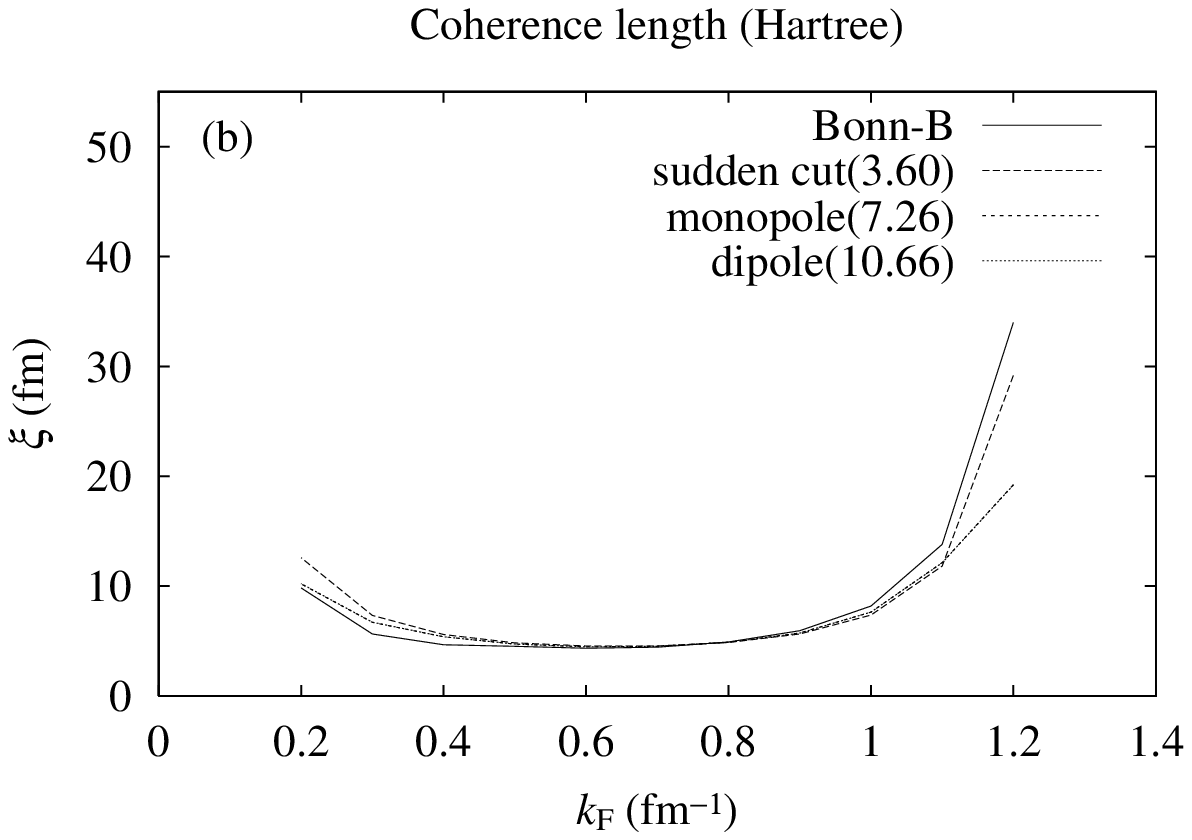}
\end{center}
\caption{
(a) Pairing gap at the Fermi surface, 
and (b) coherence length, as functions of the Fermi momentum 
$k_{\rm F}$, obtained by adopting the Bonn-B potential, the sudden cutoff in 
Ref.\cite{tm} 
and the two types of effective interaction constructed in this study.
These are the results of the Hartree-Bogoliubov calculation.
}
\label{kfdep}
\end{figure}

  Figure~\ref{kfdep} presents the results for $\Delta(k_{\rm F})$ and 
$\xi$ as functions of the Fermi momentum $k_{\rm F}$, obtained by using 
the cutoff parameters so determined. Both the monopole and the dipole 
cases reproduce the results 
from the Bonn-B potential very well, as the sudden cutoff case studied in 
Ref.\cite{tm}, in a wide and physically relevant density 
range, in the sense that pairing in finite nuclei occurs near the nuclear 
surface where density is lower than the saturation point~\cite{surf1,surf2,surf3}
and that the calculated range of $k_{\rm F}$ almost corresponds to that of 
the inner crust of neutron stars~\cite{nrel2}. 
This is our first conclusion. In the present method with a 
{\it density-independent} cutoff parameter, some small deviations remain: 
The overall slight 
peak shift to higher $k_{\rm F}$ in $\Delta(k_{\rm F})$ and the deviation 
in $\xi$ at the highest $k_{\rm F}$ are brought about by the systematic 
deviation in the critical density where the gap closes, between the 
calculations adopting bare interactions and those adopting phenomenological 
ones as mentioned above. 
The deviation at the lowest $k_{\rm F}$ is due to the feature that the 
present model is based on the mean-field picture for the finite-density 
system; this is a different point from the three-parameter fitting in 
Refs.\cite{bert,gar}. Actually, in such an extremely dilute system, the 
effective-range approximation for free scattering holds well~\cite{shift}. 

\begin{figure}[t]
\begin{center}
\includegraphics[width=8cm,keepaspectratio]{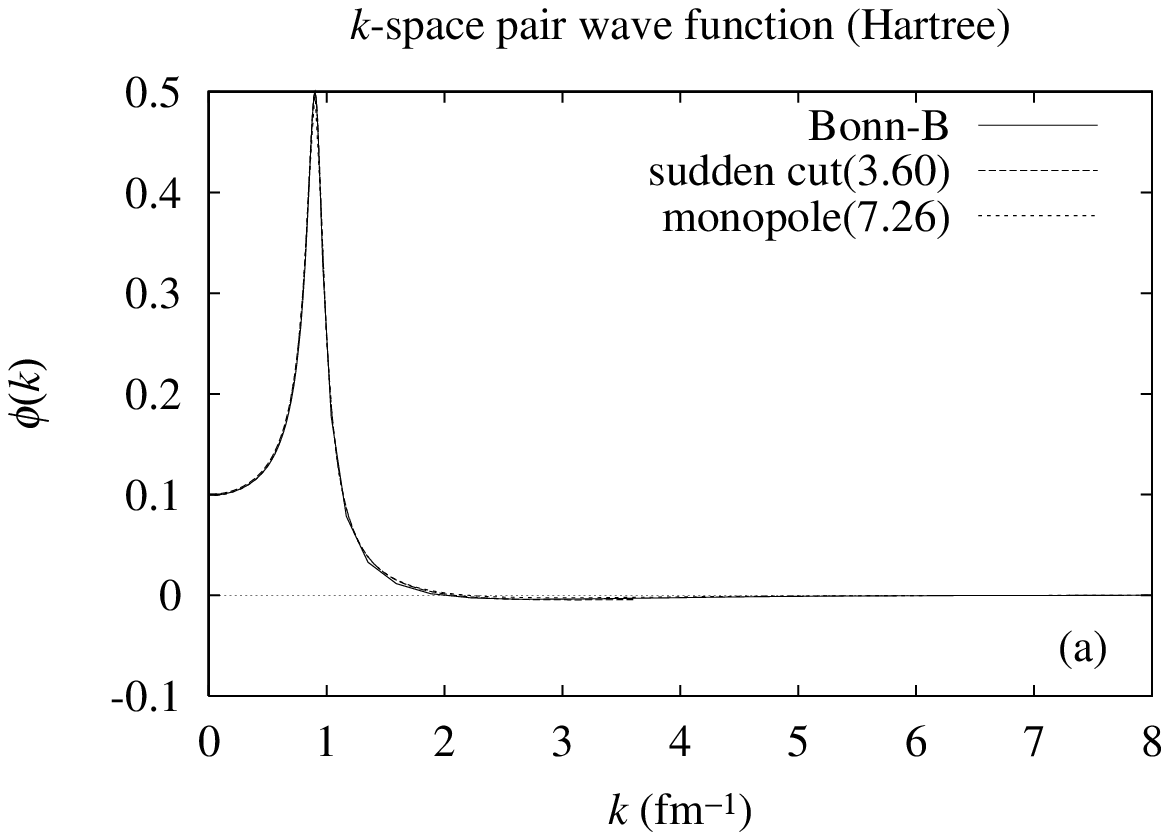}
\vskip 0.5cm
\includegraphics[width=7.5cm,keepaspectratio]{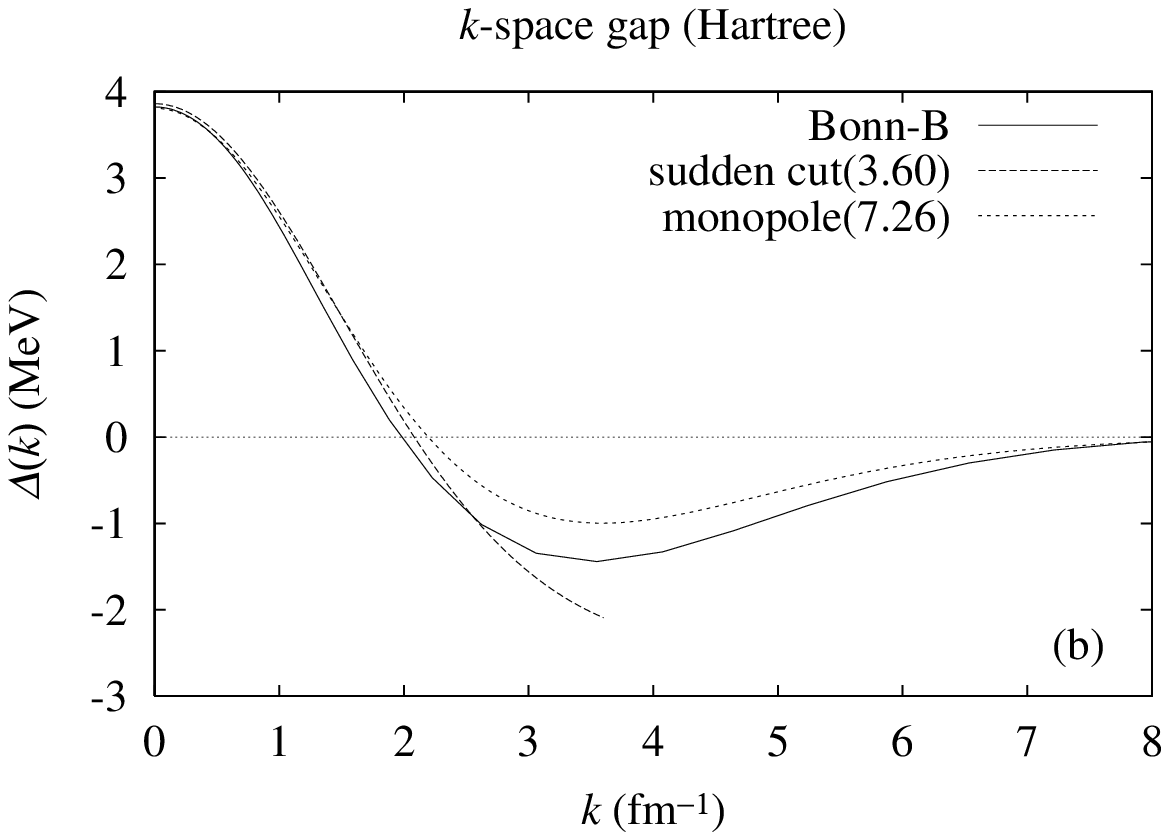}
\vskip 0.5cm
\includegraphics[width=7.5cm,keepaspectratio]{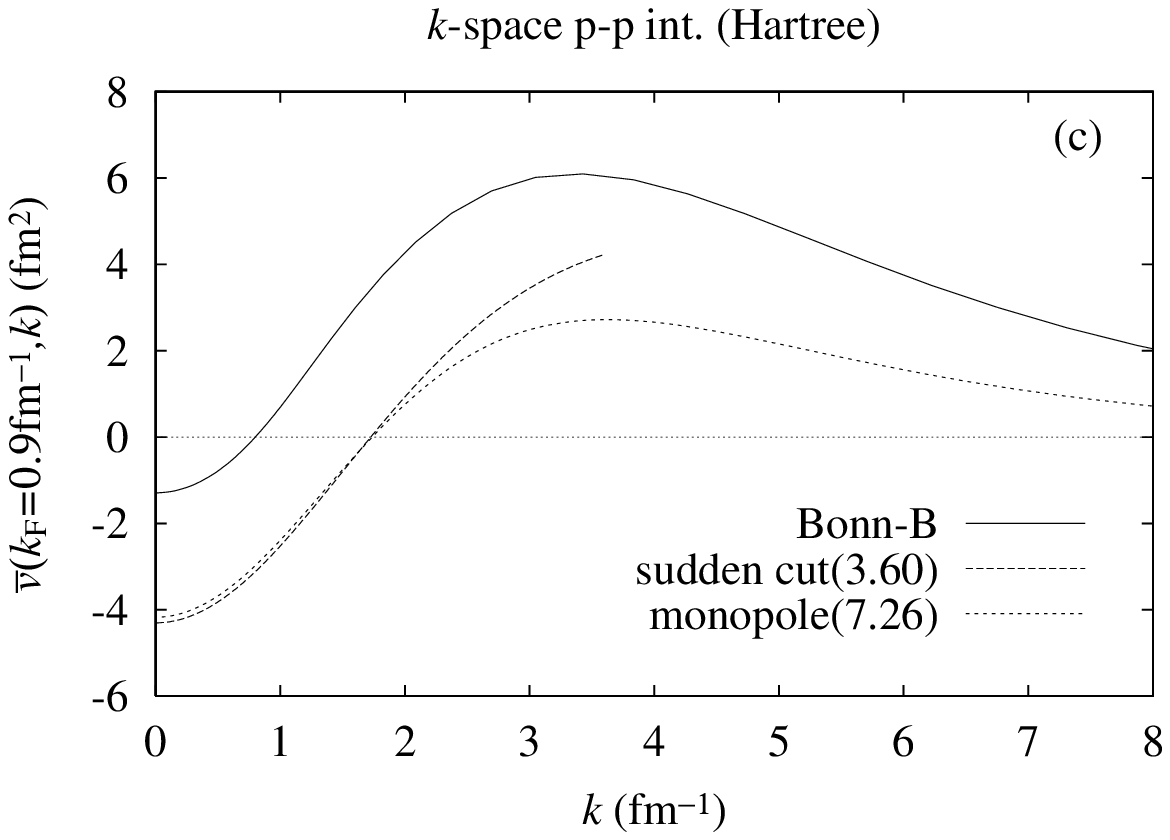}
\end{center}
\caption{
(a) Pair wave function, (b) pairing gap, and (c) 
matrix element $\bar v(k_{\rm F},k)$, as functions of the momentum $k$, 
calculated at a Fermi momentum $k_{\rm F}=~$ 0.9 fm$^{-1}$, by adopting the 
Bonn-B potential, the sudden cutoff in Ref.\cite{tm} and the 
effective interaction involving the optimal monopole form factor 
constructed in this study. 
These are the results of the Hartree-Bogoliubov calculation.
}
\label{kdep}
\end{figure}

  Next we look into the momentum dependence at $k_{\rm F}=$ 0.9 fm$^{-1}$,
where $\Delta(k_{\rm F})$ becomes almost maximum. Figure~\ref{kdep}\,(a) 
shows $\phi(k)$. 
It is evident that the monopole form factor gives the result identical to 
the Bonn-potential case as in the sudden cutoff case. This demonstrates 
clearly the effectiveness of the interaction constructed here. 
Since we confirmed that the results of the monopole and the dipole cases 
coincide within the width of the line, hereafter only the monopole case 
will be shown as a representative. This quantity peaks at $k=k_{\rm F}$ 
as seen from Eq.(\ref{pairwf}). The width of the peak represents the 
reciprocal of the coherence length. Equation (\ref{pairwf}) shows that $\phi(k)$ 
is composed of $\Delta(k)$ and the quasiparticle energy $E_{\rm qp}(k)$. 
Figure~\ref{kdep}\,(b) graphs the former. The gaps of both the sudden cutoff 
and the monopole form factor cases are almost identical up to 
$k\sim 2k_{\rm F}$, and deviations 
are seen only at larger momenta where $E_{\rm qp}(k)$ are large and 
accordingly pairing is not important. This is not a trivial result 
since the bare interaction is more repulsive than the phenomenological ones 
constructed here even at the momentum region where $\Delta(k)$ are 
almost identical as shown in Fig.~\ref{kdep}\,(c).
The reason why we compare the constructed {\it p-p channel} interaction with the 
bare interaction is as follows: Although evidently the RMF interaction 
corresponds to a medium-renormalized one, not to a bare one, here we aim at 
constructing an interaction 
similar to the Gogny force in the sense that it reproduces the pairing 
properties given by the bare interactions in spite of the fact that 
it was constructed for the finite-density system from the beginning. 
Accordingly we compare them to see the difference between an effective 
and a bare interactions which give similar pairing properties. 
This figure indicates that the difference is roughly $k$-independent. 

\begin{figure}[t]
\begin{center}
\includegraphics[width=11.27cm,keepaspectratio]{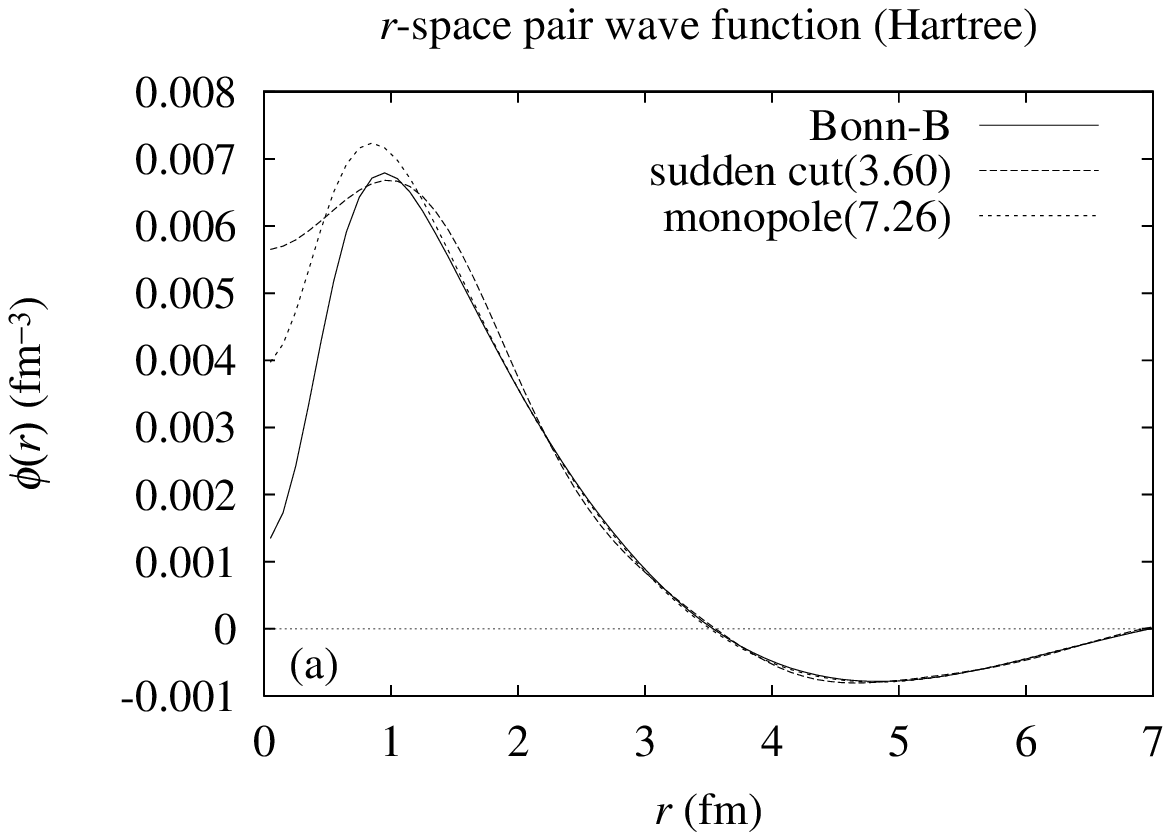}
\vskip 0.5cm
\includegraphics[width=10cm,keepaspectratio]{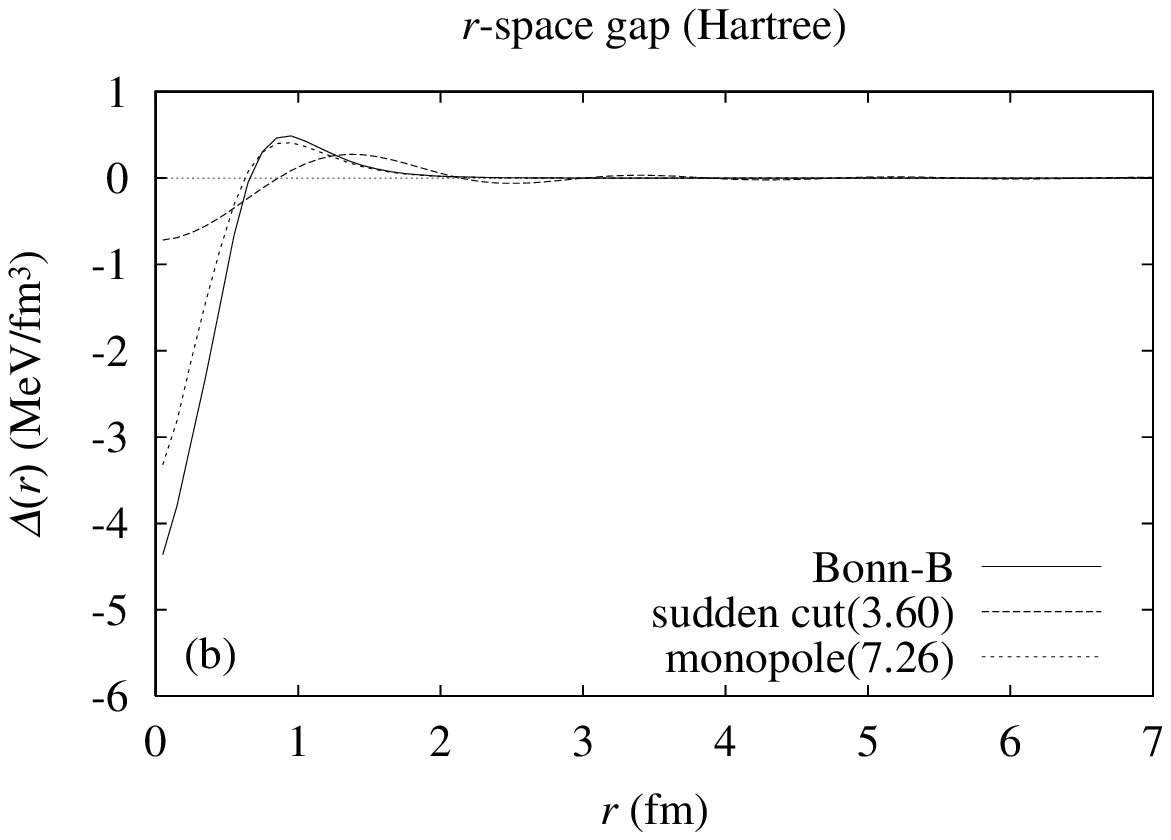}
\end{center}
\caption{
(a) Pair wave function, and (b) pairing gap, as functions of the distance $r$, 
calculated at a Fermi momentum $k_{\rm F}=~$ 0.9 fm$^{-1}$, by adopting the 
Bonn-B potential, the sudden cutoff in Ref.\cite{tm} and the 
effective interaction involving the optimal monopole form factor 
constructed in this study.
These are the results of the Hartree-Bogoliubov calculation.
}
\label{rdep}
\end{figure}

  Here we turn to the dependence on $r$, the distance between the two 
nucleons that form a Cooper pair, in order to look into the physical contents 
further. The gap equation, before the angle integration that results in 
Eq. (\ref{gapeq}), can be Fourier-transformed to 
the local form, $\Delta({\bf r})=-\bar v({\bf r})\phi({\bf r})$ 
in $r$-space in the non-relativistic limit~\cite{cor1}. 
One can see from this expression that, assuming 
$\Delta({\bf r})$ is finite, $\phi({\bf r})$ is pushed outwards when 
$\bar v({\bf r})$ has a repulsive core, as the Brueckner wave 
functions~\cite{baldo,nrel4}. This is related to an observation that the 
gap equation reduces to a Schr\" odinger equation for the relative motion 
of the two particle that form a Cooper pair in the limit of 
$v_k^2\rightarrow 0$, i.e., at high $k$\cite{shr1,shr2,gar}. 
This, on the other hand, masks practically the differences in the repulsive 
interactions at short range, in other words, widely spread in $k$-space. 
The $r$-space pair wave functions,
\begin{equation}
\phi(r)=\frac{1}{2\pi^2}\int_0^\infty\phi(k)j_0(kr)k^2dk\, ,
\end{equation}
where $j_0(kr)$ is a spherical Bessel function, 
at $k_{\rm F}=$ 0.9 fm$^{-1}$ are shown in Fig.~\ref{rdep}\,(a). 
Appreciable differences are seen only in the core region as mentioned above. 
The coherence 
length, that is a typical spatial scale of pairing correlation, is about 
6 fm at this $k_{\rm F}$ as shown in Fig.~\ref{kfdep}\,(b); 
this is almost one order of magnitude 
larger than the size of the core region. Therefore, practically we can 
safely use the p-p channel interaction, including the sudden cutoff one, 
constructed here for the gap equation.
Figure~\ref{rdep}\,(b) shows the corresponding $\Delta(r)$. 
The gaps are positive at the outside of the core and negative inside in 
all cases. Note here that the gap equation is invariant with respect 
to the overall sign inversion; we defined as $\Delta(k_{\rm F})>0$. 
Their depths at the inside region reflect the heights of the repulsive 
core. In the sudden cutoff case, $\Delta(r)$ behaves somewhat 
differently from others due to the lack of high-momentum components. 

\section{Relativistic Hartree-Fock-Bogoliubov calculation with a cutoff}

  In the preceding sections, we have shown that we can construct 
phenomenologically relativistic p-p channel interactions 
which give realistic pairing properties by introducing a {\it density-independent} 
momentum-cutoff parameter to the (no-sea) Relativistic Hartree model. 
To do this, we have fully utilized the property that only the ${\bf q}=0$ 
part of the interaction contributes to the Hartree mean field. 
To see the further applicability of the method presented above, the Relativistic 
Hartree-Fock (RHF) model, in which the ${\bf q}\neq0$ part of the interaction 
also contributes to the p-h channel, should be examined. 
In the following, we investigate the Relativistic Hartree-Fock-Bogoliubov (RHFB) 
model with a momentum-cutoff form factor, by comparison with the 
corresponding Hartree-Bogoliubov calculation. 
An RHFB calculation with a sudden cutoff was previously done by Guimar\~aes 
et al.\cite{rel1}. They obtained very large pairing gaps in the no-sea 
approximation (Fig.~4 of Ref.\cite{rel1}). 
Before studying the effects of the form factor to modulate smoothly the 
high-momentum interaction, we examine the sudden cutoff --- this is a 
straightforward extension of our previous calculation in Ref.\cite{tm} to the 
RHFB. 

\begin{figure}[t]
\begin{center}
\includegraphics[width=10cm,keepaspectratio]{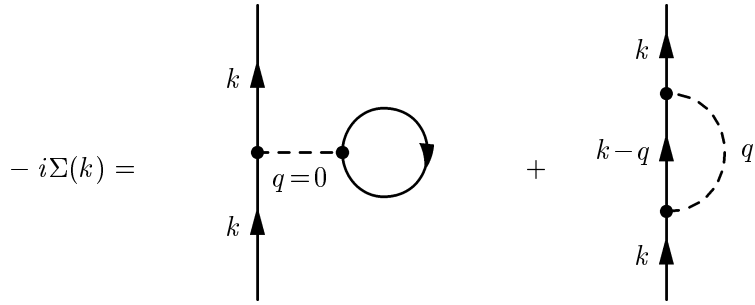}
\end{center}
\caption{
Feynman diagram representing the nucleon selfenergy in the 
Hartree-Fock model.
}
\label{f5}
\end{figure}

\begin{figure}[t]
\begin{center}
\includegraphics[width=10cm,keepaspectratio]{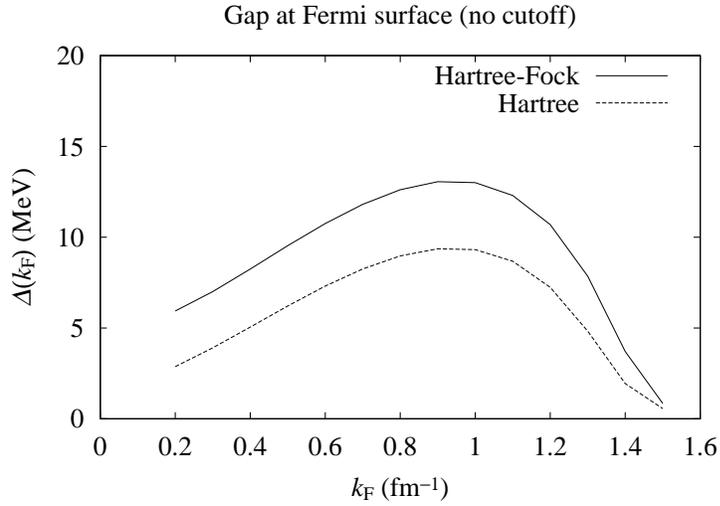}
\end{center}
\caption{
Pairing gap at the Fermi surface as a function of the Fermi momentum 
$k_{\rm F}$, obtained by the Hartree-Fock-Bogoliubov and the 
Hartree-Bogoliubov calculations with a large enough momentum cutoff, 
20 fm$^{-1}$.
}
\label{f6}
\end{figure}

  The RHF model was described in detail in Refs.\cite{sw,jam,bou}, 
for example. The difference from the Hartree model is the 
second term in Fig.~\ref{f5}. This introduces the space component of the 
vector selfenergy, $\Sigma^v$, and all the Lorentz components of $\Sigma$ 
--- the scalar $\Sigma^s$, the vector (time) $\Sigma^0$, and the vector (space) 
$\Sigma^v$ --- become momentum dependent:
\begin{equation}
\Sigma(p)=\Sigma^s(p)-\gamma^0\Sigma^0(p)+{\bm \gamma}\cdot{\bf p}\Sigma^v(p)\, .
\end{equation}
In contrast, in the Hartree model, $\Sigma^v$ is not present, $\Sigma^s$ and 
$\Sigma^0$ are momentum independent, and $\Sigma^0$ is fully determined by the 
input density. Their concrete forms are
\begin{eqnarray}
\Sigma^s(p)&&=-\frac{\gamma}{(2\pi)^3}\frac{g_\sigma^2}{m_\sigma^2}
\int_0^{k_{\rm F}}d^3k\frac{M^\ast(k)}{E^\ast(k)}
\nonumber \\
&&+\frac{1}{4\pi^2k}\int_0^{k_{\rm F}}dkk\frac{M^\ast(k)}{E^\ast(k)}
\Bigl[\frac{1}{4}g_\sigma^2\Theta_\sigma(p,k)-g_\omega^2\Theta_\omega(p,k)\Bigr]\, ,
\nonumber \\
\Sigma^0(p)&&=-\frac{\gamma}{(2\pi)^3}\frac{g_\omega^2}{m_\omega^2}
\int_0^{k_{\rm F}}d^3k
\nonumber \\
&&-\frac{1}{4\pi^2k}\int_0^{k_{\rm F}}dkk
\Bigl[\frac{1}{4}g_\sigma^2\Theta_\sigma(p,k)
     +\frac{1}{2}g_\omega^2\Theta_\omega(p,k)\Bigr]\, ,
\nonumber \\
\Sigma^v(p)&&=-\frac{1}{4\pi^2k^2}\int_0^{k_{\rm F}}dkk\frac{k^\ast}{E^\ast(k)}
\Bigl[\frac{1}{2}g_\sigma^2\Phi_\sigma(p,k)+g_\omega^2\Phi_\omega(p,k)\Bigr]\, ,
\end{eqnarray}
with 
\begin{eqnarray}
&&\Theta_i(p,k)=\ln\left|\frac{A_i(p,k)+2pk}{A_i(p,k)-2pk}\right|\, ,
\nonumber \\
&&\Phi_i(p,k)=\frac{1}{4pk}A_i(p,k)\Theta_i(p,k)-1\, ,
\nonumber \\
&&A_i(p,k)={\bf p}^2+{\bf k}^2+m_i^2\, , \quad\quad\quad\quad(i=\sigma, \omega)\, ,
\end{eqnarray}
as Eqs. (5.76) - (5.79) in Ref.\cite{sw}. The retardation effect is neglected 
here. Note that the momentum dependence of 
$M^{\ast}$ is stemming from that of $\Sigma^s$: $M^{\ast}=M+\Sigma^s$ 
(see Eq.(\ref{effmass})). 
When pairing is introduced, $\int_0^{k_{\rm F}}$ is replaced by 
$\int_0^\infty v_k^2$, and therefore $\Sigma^s(p)$, $\Sigma^0(p)$, $\Sigma^v(p)$, 
and $\Delta(p)$ (Eq.(\ref{gapeq})) couple through Eqs.(\ref{v2}) and (\ref{spe}). 
When the momentum space is discretized to $n$ meshes, these quantities form 
a set of $4n$-dimensional coupled non-linear equations; this is a contrast to the 
$n+1$-dimensional ones --- $\Delta(p)$ and a momentum-independent $M^{\ast}$ or 
$\Sigma^s$ --- in the Hartree case. 

  In order to reproduce the saturation, the coupling constants are adjusted to 
$g_\sigma^2=$ 83.11, $g_\omega^2=$ 108.05 with the masses being unchanged 
\cite{sw}. We adopt this parameter set for the time being. Other sets will be 
examined later. Note that the pairing contribution to the energy density 
around the saturation point is negligible. First we compare the gap at the 
Fermi surface calculated without a cutoff in RHFB and RHB, 
in Fig.~\ref{f6}. We call them the full calculations hereafter. 
The calculated gaps in the former are larger than those in the latter, by about 
4 MeV at the maximum, for example. This difference is brought about by that in 
the coupling constants rather than the Fock effect itself \cite{carl}. 
Actually we confirmed that a Hartree calculation with the coupling constants 
of the Hartree-Fock gave almost the same gaps as those given by the Hartree-Fock 
calculation although such a calculation destroys the saturation completely. 
This result reflects the property that the gap is not sensitive to the detail 
of the single-particle states. If the coupling constants of Ref.\cite{rel1}, 
$g_\sigma^2=$ 96.392, $g_\omega^2=$ 129.260 are adopted, calculated gaps become 
even larger; about 15 MeV at the maximum, for example. But it is evident that 
this is still smaller than their calculated value in their Fig.~4. The reason 
for this difference is not clear. 

\begin{figure}[t]
\begin{center}
\includegraphics[width=10cm,keepaspectratio]{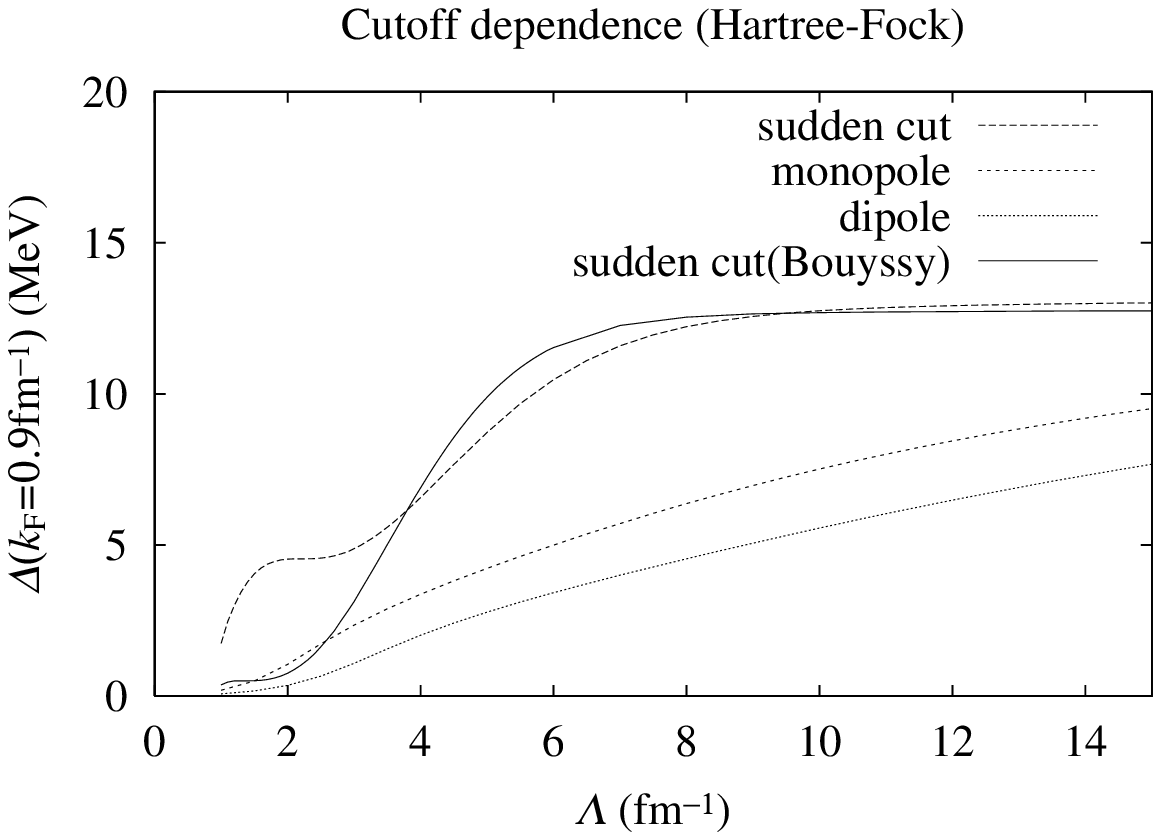}
\end{center}
\caption{
The same as Fig.1(b) but for the Hartree-Fock-Bogoliubov calculations.
}
\label{f7}
\end{figure}

  Now let us introduce a sudden cutoff in the upper bound of the momentum 
integrations in $\Sigma(p)$'s and $\Delta(p)$ as in the previous Hartree case. 
The result is graphed in Fig.~\ref{f7} by the long-dashed curve. 
Although a plateau appears around $\Lambda=$ 2 fm$^{-1}$ as in the 
corresponding case in Fig.~\ref{cutdep}\,(b), the gap value of the present 
plateau is about 4.5 MeV, which is larger than the physical value given by the 
bare interaction, about 2.8 MeV at this density.  Applying the same procedure 
as in the previous Hartree case, the obtained optimal cutoff is 1.26 fm$^{-1}$. 
This indicates that the attractive interaction alone accounts for the physical 
magnitude of the gap since the $\Lambda$ of the plateau corresponds to the $k$ 
at which $\bar v(k_{\rm F},k)$ turns from attractive to repulsive~\cite{tm}. 
Note that evidently the RHFB calculation with this sudden cutoff can not be 
applied to the case with $k_{\rm F}>\Lambda$. 

\begin{figure}[t]
\begin{center}
\includegraphics[width=10cm,keepaspectratio]{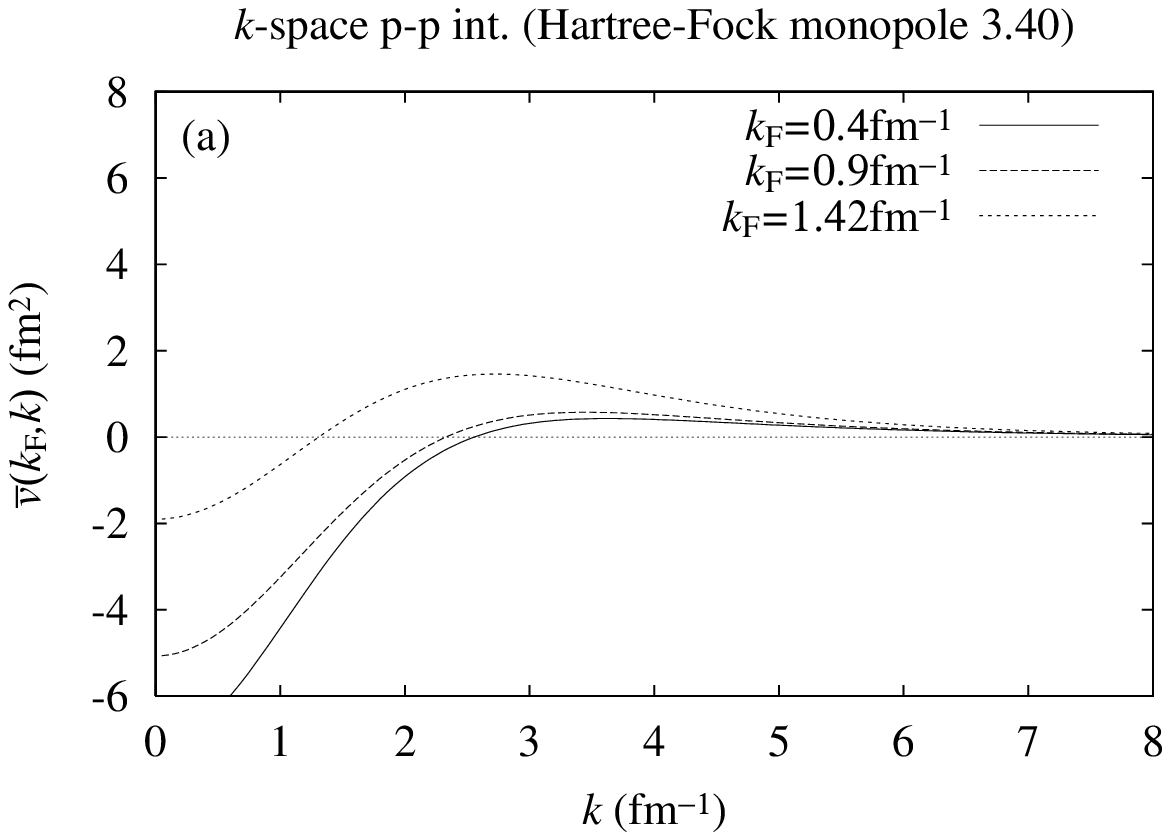}
\vskip 0.5cm
\includegraphics[width=10cm,keepaspectratio]{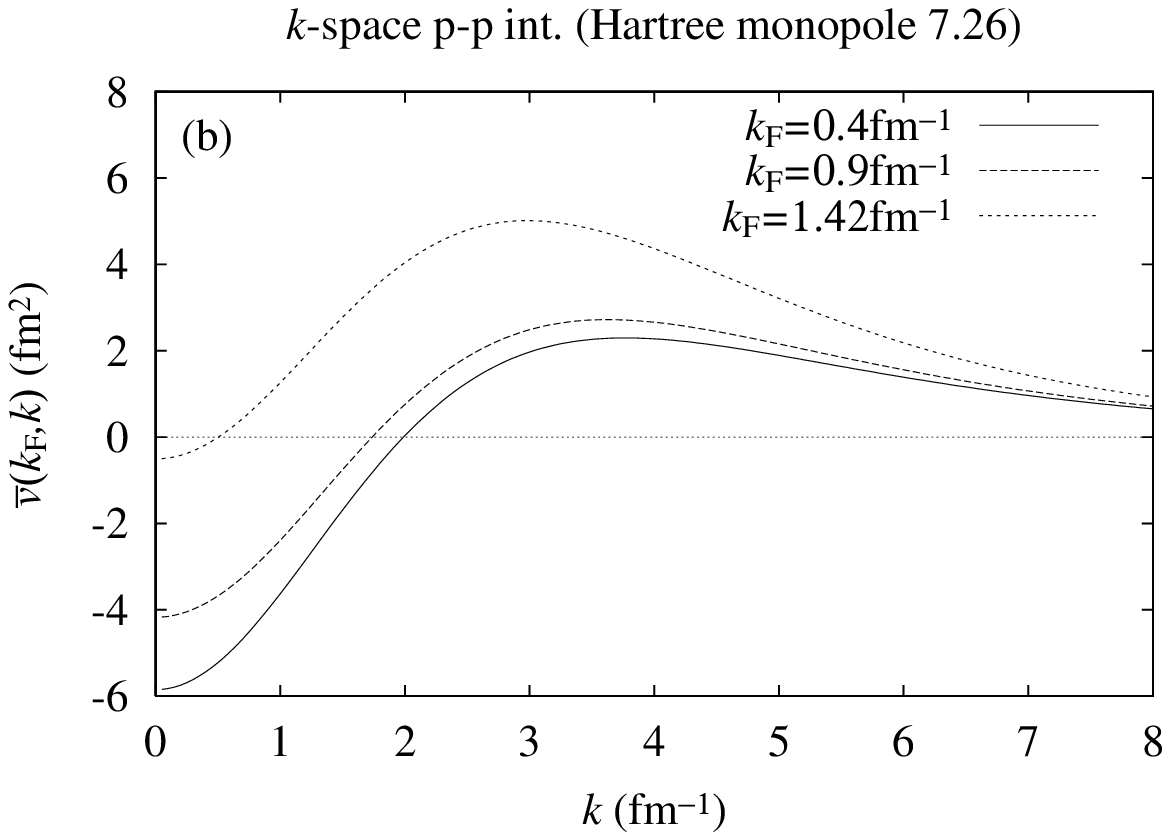}
\end{center}
\caption{
Matrix element $\bar v(k_{\rm F},k)$, as a function of the momentum $k$, 
obtained by (a) the Hartree-Fock, and (b) the Hartree-Fock-Bogoliubov 
calculations with the optimal monopole form factors, 
calculated at three Fermi momenta $k_{\rm F}$.
}
\label{f8}
\end{figure}

\begin{figure}[t]
\begin{center}
\includegraphics[width=9.76cm,keepaspectratio]{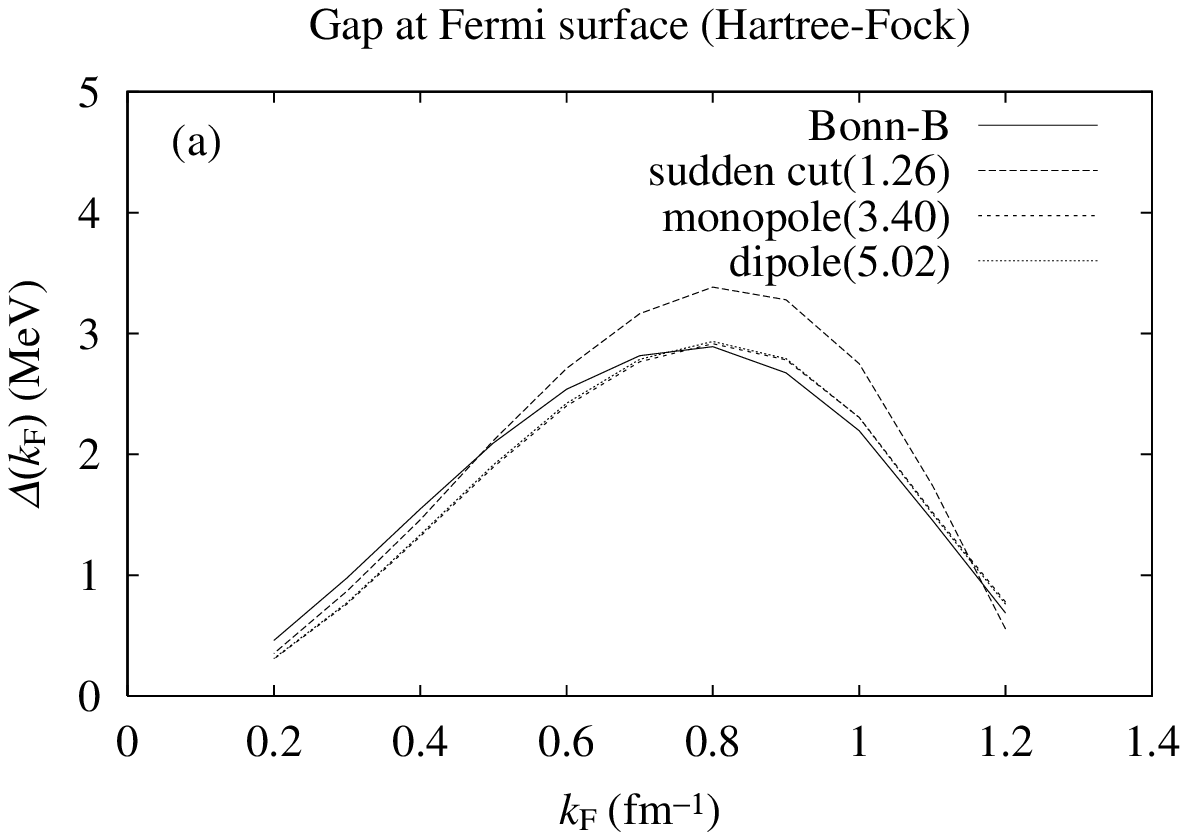}
\vskip 0.5cm
\includegraphics[width=10cm,keepaspectratio]{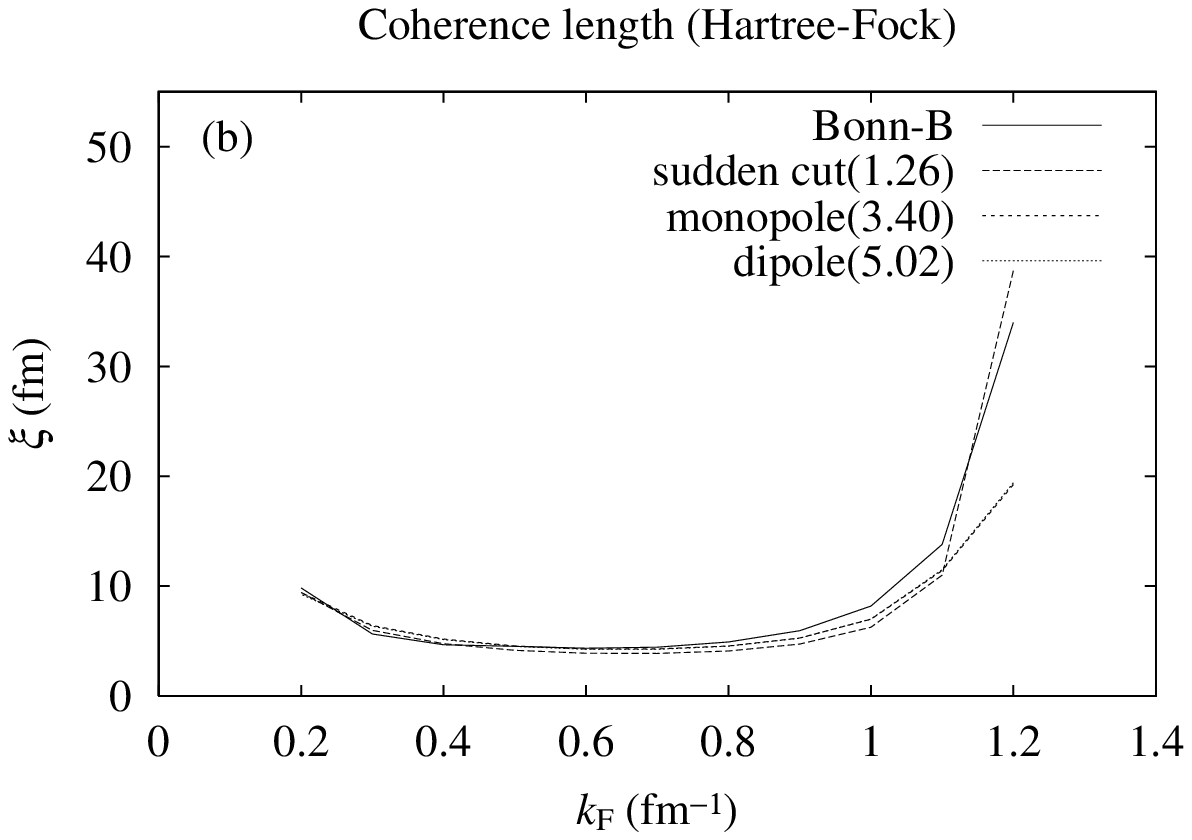}
\end{center}
\caption{
The same as Fig.2 but for the Hartree-Fock-Bogoliubov calculations.
}
\label{f9}
\end{figure}

  Next we proceed to the smooth cutoff --- the form factor in the nucleon-meson 
vertices. The monopole and the dipole types are examined here. Their 
$\Lambda$-dependence is included in Fig.~\ref{f7}. The optimal cutoffs are 
$\Lambda=$ 3.40 fm$^{-1}$ and 5.02 fm$^{-1}$ for the monopole and the dipole 
types, respectively. These values are smaller than the corresponding ones in the 
Hartree cases since the original gap values given by the full calculation 
are larger. Interestingly, however, their ratio, 
$\Lambda({\rm monopole})/\Lambda({\rm dipole})$, almost 
coincides with that of the Hartree case. The p-p channel 
interactions including the optimal monopole form factor are compared in 
Fig.~\ref{f8}. The high-momentum repulsive part, in particular at lower 
densities, is strongly suppressed in the RHFB case. This testifies the discussion 
about Fig.~\ref{f7} above that the optimal cutoffs are smaller. 
In Fig.~\ref{f9} we present $\Delta(k_{\rm F})$ and $\xi$ calculated by 
adopting the optimal cutoffs. They reproduce the values given by the Bonn-B 
potential to an extent similar to the Hartree case or a little better except 
$\Delta(k_{\rm F})$ of the sudden cutoff case. Again the results of the monopole 
and the dipole form factors coincide with each other within the width of 
the line. 

\begin{figure}[t]
\begin{center}
\includegraphics[width=10.32cm,keepaspectratio]{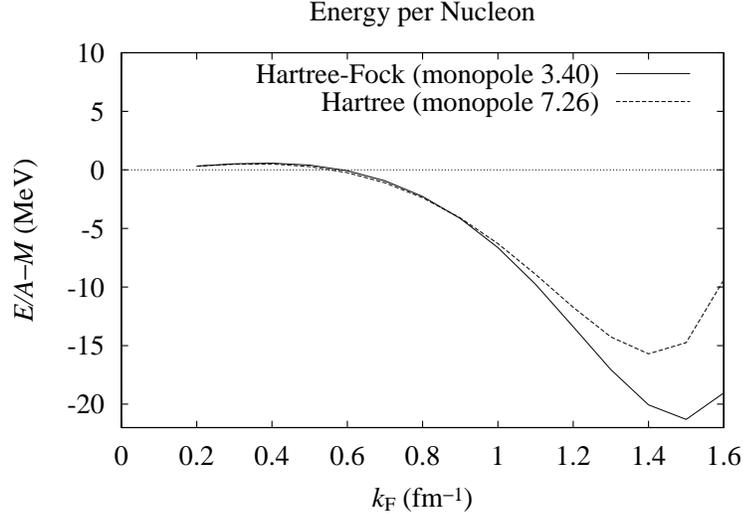}
\end{center}
\caption{
Energy density including the pairing contributions (both in the 
Hartree-Fock-Bogoliubov and the Hartree-Bogoliubov calculations) 
and the cutoff effects (only in the former) as a function of the 
Fermi momentum $k_{\rm F}$.
}
\label{f10}
\end{figure}

\begin{figure}[t]
\begin{center}
\includegraphics[width=10.32cm,keepaspectratio]{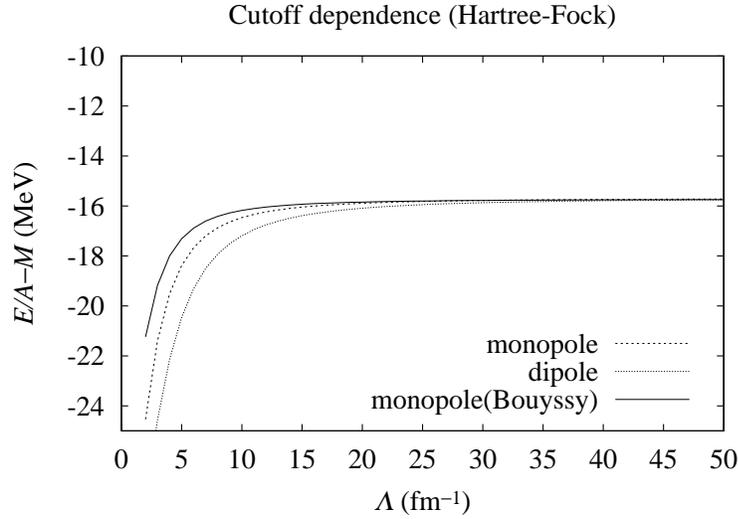}
\end{center}
\caption{
Energy density at the saturation density corresponding to 
$k_{\rm F}^0$ obtained by the Hartree-Fock calculations 
(without pairing), calculated as a function of the 
momentum-cutoff parameter in the form factors. 
$k_{\rm F}^0=$ 1.42 fm$^{-1}$ for the first two cases, while 
1.30 fm$^{-1}$ for the third case. 
Note that the scale of the abscissa is different from Fig.7.
}
\label{f11}
\end{figure}

  Although the pairing properties of the RHFB with the optimal form factor 
presented in Fig.~\ref{f9} are essentially the same as those of the 
corresponding Hartree calculation, an essential feature of the Hartree-Fock 
model is that the form factor can affect the saturation property. 
Therefore we have to check this before concluding the applicability of the 
present procedure. The energy density is graphed as a function of the 
Fermi momentum in Fig.~\ref{f10}. This figure can be compared with Fig.~40 
in Ref.\cite{sw}. Here the pairing correlation energy (both cases) and the 
cutoff contribution (Hartree-Fock case only) are additionally included. 
Although these two curves almost coincide with each other at low densities, 
the cutoff effect destroys the saturation in the RHFB case. This is because 
suppressing the 
high-momentum repulsion breaks the balance between the attraction and the 
repulsion, and the latter contribute more at higher densities (see 
Fig.~\ref{f8}\,(a)). To see this more closely, the cutoff dependence of the 
energy density without the pairing contribution at the saturation density, 
$k_{\rm F}^0=$ 1.42 fm$^{-1}$, is shown in Fig.~\ref{f11}. In the 
large-$\Lambda$ limit, the energy densities approaches to about -15.69 MeV, 
which is 0.38 \% less bound than the original value in Ref.\cite{sw}, 
-15.75 MeV, because of the instantaneous approximation. This figure shows that 
\begin{equation}
\Lambda_{\rm p-p}<\Lambda_{\rm p-h}\, ,
\end{equation}
if we call the cutoff which reproduces the pairing properties and which does 
the saturation properties $\Lambda_{\rm p-p}$ and $\Lambda_{\rm p-h}$, 
respectively.

  Up to now, we used a parameter set of Serot and Walecka. To complete the 
discussion, other sets with different characteristics are also examined here. 
One is a set used by Bouyssy et al., $g_\sigma^2=$ 69.62, $g_\omega^2=$ 153.81, 
$m_\sigma=$ 440 MeV, $m_\omega=$ 783 MeV, and $M=$ 938.9 MeV, which gives 
saturation at $k_{\rm F}^0=$ 1.30 fm$^{-1}$\cite{bou}. A distinct feature of 
this set is that the plateau appears at very small pairing gap (see 
Fig.~\ref{f7}). This indicates that the repulsive part also contributes to the 
gap as in the Hartree case, and accordingly, $\Lambda_{\rm p-p}$ will be 
larger than in the case of Serot and Walecka's set. Actually, the optimal 
monopole cutoff is 5.22 fm$^{-1}$, which is 
larger than 3.40 fm$^{-1}$ in the previous case. But Fig.~\ref{f11} shows that 
still $\Lambda_{\rm p-p}<\Lambda_{\rm p-h}$. Another is a set used by 
Jaminon et al.\cite{jam}. An interesting feature of this set is that the 
saturation (at $k_{\rm F}^0=$ 1.36 fm$^{-1}$) is given by introducing a ``weak" 
form factor~\cite{bro},
\begin{equation}
f(q^2)=\sqrt{\frac{\Lambda^2}{\Lambda^2-q^2}}\, ,
\end{equation}
from the beginning.
Note that here $q^2$ is the square of the 4-momentum transfer and therefore 
this contains a retardation effect. The parameters are $g_\sigma^2=$ 93.87, 
$g_\omega^2=$ 127.55, $m_\sigma=$ 550 MeV, $m_\omega=$ 782.8 MeV, and 
$\Lambda=$ 7.754 fm$^{-1}$. Since $M$ is not shown explicitly, 939 MeV 
is adopted. The RHFB calculation with this form factor under the instantaneous 
approximation gives pairing gaps which is larger than the physical values; 
$\Delta(k_{\rm F})\simeq$ 8.5 MeV at the maximum, for example. 
This means that still $\Lambda_{\rm p-p}<\Lambda_{\rm p-h}$ is necessary to 
reproduce the physical pairing gap. 

  In all the calculations above, the cutoff in the form factors is common to 
$\sigma$ and $\omega$. Referring to the Bonn potential and Ref.\cite{hi2}, 
for example, $\Lambda_\sigma\neq\Lambda_\omega$ is another option. 
Of course large deviations from $\Lambda_\sigma=\Lambda_\omega$ destroy 
the saturation at least when the coupling constants are kept unchanged. 
We examined some cases in a limited range, 
$\frac{\Lambda_\omega}{\Lambda_\sigma}=$ 
0.70 $(\simeq\frac{m_\sigma}{m_\omega})$ - 1.2, but the results were negative. 

\section{Conclusions}

  We have constructed phenomenologically a particle-particle channel 
interaction which suits the gap equation for nuclear matter. 
This was done by introducing a density-independent 
momentum-cutoff parameter to the one-boson exchange interaction derived 
from the Lagrangian of the RMF model and adjusting it to the pairing properties 
obtained by the Bonn-B potential. The model pairing properties were 
calculated by using the RMF model in the p-h channel and the Bonn-B potential 
in the p-p channel utilizing the properties that the RMF model simulates 
crudely\footnote
{
The coupling constants must be density-dependent in order to simulate 
the $G$-matrices quantitatively.
} 
the $G$-matrices in the DBHF calculation and that the pairing properties 
are not sensitive to the detail of the single-particle states. 
By this procedure we aimed at constructing an interaction like the Gogny 
force in the sense that it reproduces the pairing 
properties given by the bare interactions in spite of the fact that 
it was constructed for the finite-density system from the beginning. 
The actual determination of the optimal cutoff parameter was done by the 
method proposed in our preceding letter for the Hartree mean field plus 
a sudden cutoff~\cite{tm}. 

  In the present paper, first we applied this method to the (no-sea) 
Relativistic Hartree model plus various types of form factor which 
modulates the high-momentum repulsion that spoils the pairing properties. 
Among the four types examined, the monopole and the dipole form factors 
exhibit desired properties. Close analyses in momentum space and coordinate 
space have clarified that the gap equation involves a mechanism to mask 
the difference in the short-range repulsion between the bare and the 
in-medium effective interactions and that, in the typical spatial scale 
of pairing phenomena determined by the coherence length, practically there are 
no difference in the pairing properties. 

  Second we performed RHFB calculations with and without a momentum cutoff 
--- a sudden cutoff and two types of form factor. The reason why we examined 
the RHFB separately is that the form factor which modulates the 
${\bf q}\neq0$ part of the interaction also affects the p-h channel. 
The same procedure was applied; the resulting optimal cutoffs are smaller 
than those for the Hartree model because the original gap values given by the 
full calculation without a cutoff are larger. Interestingly, the ratio
$\Lambda({\rm monopole})/\Lambda({\rm dipole})$ almost coincides with 
that for the Hartree model. Having confirmed that these optimal form factors 
reproduce $\Delta(k_{\rm F})$ and $\xi$ with a quality similar to that in the 
Hartree calculation, we looked at the saturation curve of the energy density. 
Although the results of the RHB and the RHFB each with the optimal form factor 
coincide with each other up to $k_{\rm F}\sim$ 1 fm$^{-1}$, the result for 
the latter starts to deviate at larger $k_{\rm F}$. This indicates that 
$\Lambda_{\rm p-p}<\Lambda_{\rm p-h}$ is necessary in order also to reproduce 
the saturation simultaneously using the RMF interaction both in the p-h and 
the p-p channels. This holds also for the other parameter sets with 
characteristic features mentioned in the previous section. 

  These results support the observation that the high-momentum components 
of the original RMF interaction should be refined. 
The modulated interaction obtained here by the proposed one-parameter fitting 
can be successfully used for the Relativistic Hartree-Bogoliubov calculation. 
But its applicability to the Relativistic Hartree-Fock-Bogoliubov 
calculation is limited to low densities.

\newpage
\noindent
Table 1

\noindent
Classification of various relativistic approaches to nucleon-nucleon pairing 
\begin{center}
\begin{tabular}{llll} \hline
Type       & p-h channel & p-p channel(lowest order) & References         \\ \hline
B          & DBHF        & bare                      & \cite{rel3}               \\
B$^\prime$ & DBHF        & effective range           & \cite{shift}              \\
P1         & RMF         & RMF                       & \cite{kr,rel1,rel2,mm,tm} \\
P2a        & RMF         & another phenomenological  & \cite{kr,rr}              \\
P2b        & RMF         & bare                      & \cite{rr,mt}              \\
P2b$^\prime$ & RMF       & effective range           & --                 \\ \hline
\end{tabular}
\end{center}

\newpage

\newpage

\newpage

\newpage

\newpage

\newpage

\newpage

\newpage

\newpage

\newpage

\newpage

\end{document}